# Gigahertz repetition rate, sub-femtosecond timing jitter optical pulse train directly generated from a mode-locked Yb:KYW laser

Heewon Yang,[1] Hyoji Kim,[1] Junho Shin,[1] Chur Kim,[1] Sun Young Choi,[2] Guang-Hoon Kim,[3] Fabian Rotermund,[2] and Jungwon Kim[1,*]

[1]School of Mechanical, Aerospace and Systems Engineering, Korea Advanced Institute of Science and Technology (KAIST), Daejeon 305-701, South Korea
[2]Department of Physics and Division of Energy Systems Research, Ajou University, Suwon 443-749, South Korea
[3]Russia Science Seoul (RSS) Center, Korea Electrotechnology Research Institute (KERI), Seoul 121-835, South Korea
*Corresponding author: jungwon.kim@kaist.ac.kr

We show that a 1.13-GHz repetition rate optical pulse train with 0.70 fs high-frequency timing jitter (integration bandwidth of 17.5 kHz – 10 MHz, where the measurement instrument-limited noise floor contributes 0.41 fs in 10 MHz bandwidth) can be directly generated from a free-running, single-mode diode-pumped Yb:KYW laser mode-locked by single-walled carbon nanotube (SWCNT)-coated mirrors. To our knowledge, this is the lowest timing jitter optical pulse train with the GHz repetition rate ever measured. If this pulse train is used for direct sampling of 565-MHz signals (Nyquist frequency of the pulse train), the demonstrated jitter level corresponds to the projected effective-number-of-bit (ENOB) of 17.8, which is much higher than the thermal noise limit of 50-ohm load resistance (~14 bits).

The availability of optical pulse trains with both gigahertz (GHz)-level repetition rate and sub-femtosecond-level timing jitter is crucial for extension of widespread applications of ultrafast mode-locked lasers in telecommunication and information systems, such as high-speed, high-resolution photonic analog-to-digital converters (ADCs) [1-4] and pulsed laser-based optical interconnects [5]. In recent years, there have been dramatic progresses in lowering the timing jitter in mode-locked solid-state and fiber lasers: currently, various types of mode-locked solid-state and fiber lasers can generate optical pulse trains with well below a femtosecond timing jitter [6-12]. However, all these sub-femtosecond timing jitter performances have been demonstrated at ~100 MHz repetition rate.

To increase the repetition rate to the GHz regime with low timing jitter, several approaches have been demonstrated. First, an external repetition-rate multiplier (such as a Fabry-Perot cavity) locked to the laser oscillator can be employed. For example, a 2-GHz repetition rate optical pulse train with timing jitter of ~27 fs (1 kHz – 10 MHz) was achieved from a 200-MHz fiber laser [13] and was further used in a photonic ADC [3]. Second, passive or hybrid harmonic mode-locking can be used. Although the residual jitter between the driving microwave source and laser output can be in the few-fs regime [14], the absolute timing jitter has been often limited to a few ps [15-17]. Third, semiconductor mode-locked lasers can naturally achieve very high repetition rates with multi-tens of GHz. However, the achievable pulse width in the ps range is still too long and the resulting timing jitter is also limited to ~1 ps regime [18]. Finally, GHz repetition rate, fundamentally mode-locked solid-state, fiber, or waveguide lasers [19-22] can provide a simple and direct way for lower timing jitter. So far, the best jitter performance of GHz mode-locked lasers is in the ~20 fs level [19], limited by both the laser performance itself and the resolution of measurement method based on direct photodetection and microwave mixers.

In this Letter, we show that a 1.13-GHz repetition rate optical pulse train with 0.70 fs high-frequency timing jitter (integration bandwidth of 17.5 kHz – 10 MHz, where the measurement instrument-limited noise floor contributes 0.41 fs in 10-MHz bandwidth) can be directly generated from a free-running, single-mode diode-pumped Yb:KYW laser mode-locked by single-walled carbon nanotube (SWCNT) saturable absorber mirrors. To our knowledge, this is the lowest timing jitter optical pulse train with the GHz repetition rate ever measured. Note that, if this pulse train is used for direct sampling of 565-MHz signals (Nyquist frequency of the pulse train), the projected sampling jitter-limited effective-number-of-bit (ENOB) resolution is 17.8 bits, which is already much higher than the thermal noise limit of 50-ohm load resistance (~14 bits) [1]. If the same pulse train is used for the wavelength-division multiplexed photonic ADC such as in [3], the achievable jitter-limited ENOB can be significantly improved to >11 bits for 40-GHz input signal.

Figure 1 shows (a) schematic of the 1.13-GHz Yb:KYW laser, (b) measured optical spectrum, (c) radio-frequency (RF) spectrum, and (d) relative intensity noise (RIN) spectrum. The laser has a similar design and structure as previously demonstrated in [23], only with a notable difference that two SWCNT-coated mirrors (M3 and OC in Fig. 1(a)) are used as saturable absorbers to increase modulation depth and enhance the long-term stability of the laser. At the expense of better stability, the 3-dB bandwidth is decreased to 4.0 nm centered at 1037 nm, which corresponds to the transform-limited full-width half maximum (FWHM) pulse width of 282 fs (sech$^2$-shape), compared to the previous 168 fs pulse width in [23]. The output power is 35 mW from a 0.3 % output coupler, when pumped by a fiber-coupled, 750 mW, 980 nm, single-mode laser diode.

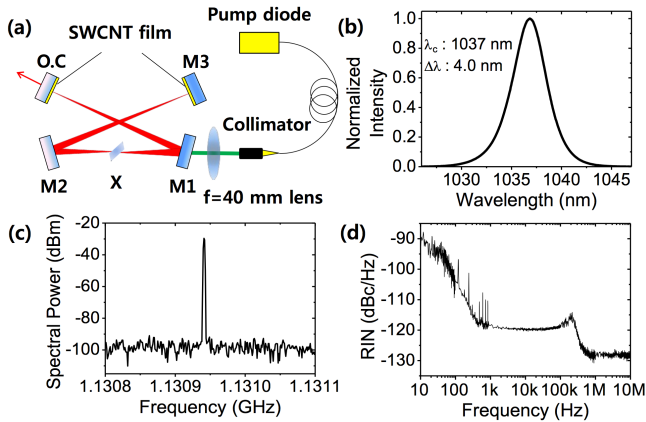

Fig. 1. (a) Schematic of the 1.13-GHz repetition rate Yb:KYW laser. M1: broadband dielectric curved mirror with radius of curvature (ROC) = 30 mm. M2: Gires-Tournois interferometer (GTI) mirror with -800 fs$^2$ dispersion at 1045 nm and ROC = 30 mm. M3: broadband dielectric flat mirror with SWCNT-SA coating. OC: 0.3 % output coupler with SWCNA-SA coating. X: 2-mm, 5%-doped, Brewster-cut Yb:KYW laser crystal. Pump diode: 980 nm, 750 mW single-mode laser diode. (b) Measured output optical spectrum. The FWHM bandwidth is 4.0 nm at 1037 nm center wavelength. (c) Measured RF spectrum with 1-kHz resolution bandwidth and 300-kHz span. (d) Measured RIN spectrum from 10 Hz to 10 MHz offset frequency (integrated RIN of 0.14 %rms).

For high-resolution timing jitter characterization of the 1.13-GHz Yb:KYW laser, a timing detector method based on balanced optical cross-correlation (BOC) [24] is employed. The experimental setup for the timing jitter characterization is shown in Fig. 2. To measure the jitter of the laser under test (LUT, 1.13-GHz Yb:KYW laser in this work), one needs a reference laser with similar or lower jitter than the LUT. In this work, as we have only one Yb:KYW laser, we used an Yb-fiber laser with ~20-as jitter performance [12] as the reference laser. Note that the repetition rate of the used Yb-fiber laser is 188 MHz, which is 6th sub-harmonic of the Yb:KYW laser repetition rate (1.13 GHz). As a result, every sixth pulse from the Yb:KYW laser is used for locking the LUT to the reference laser using the BOC. Note that the different repetition rate of the reference laser and the LUT does not have an impact on the BOC jitter measurement. Because the pulse energy directly generated from the Yb:KYW oscillator is insufficient for high-resolution timing detection using a nonlinear process in the BOC (type-II phase-matched second-harmonic generation by BBO crystals in this work), the output from the Yb:KYW laser is amplified by an external Yb-doped fiber amplifier (YDFA) as shown in Fig. 2. The output from the YDFA is ~220 mW when pumped by a 750 mW, 976 nm laser diode. After dechirping by a grating pair, the resulting pulse width and pulse energy used in the BOC are ~300 fs and 71 pJ, respectively. The pulse width and pulse energy of the Yb-fiber laser (reference laser) are ~40 fs and 0.45 nJ, respectively, in the BOC. The resulting timing detection slope from the balanced photodetector in the BOC is 0.24 mV/fs when transimpedance gain of 6 kΩ is used. Compared to the recent 188-MHz repetition rate, 20-as jitter Yb-fiber laser jitter characterization [12], the resulting timing detection sensitivity is about 100 times lower due to longer pulse width and lower output pulse energy of the Yb:KYW laser. However, as shown in Fig. 3, the resulting measurement noise floor still reaches 10$^{-8}$ fs$^2$/Hz level. This corresponds to -185 dBc/Hz equivalent phase noise floor at the 1.13 GHz carrier frequency, where direct photodetection-based measurement methods cannot reach. The error signal from the BOC is used to lock the repetition rate between the two lasers [1.13 GHz (6th harmonic of 188 MHz) versus 188 MHz] with the lowest possible locking bandwidth, which allows the jitter spectral density characterization of the free-running Yb:KYW laser outside the locking bandwidth (~17.5 kHz in this work). Note that 17.5 kHz was the minimal bandwidth we could use for getting stable repetition-rate locking between the Yb:KYW laser and the Yb-fiber laser. The required locking bandwidth is higher than the case of synchronization between two Yb-fiber lasers (where it was ~2 kHz locking bandwidth) [12], which is mainly due to higher sensitivity of a solid-state laser to the vibration and acoustic noise.

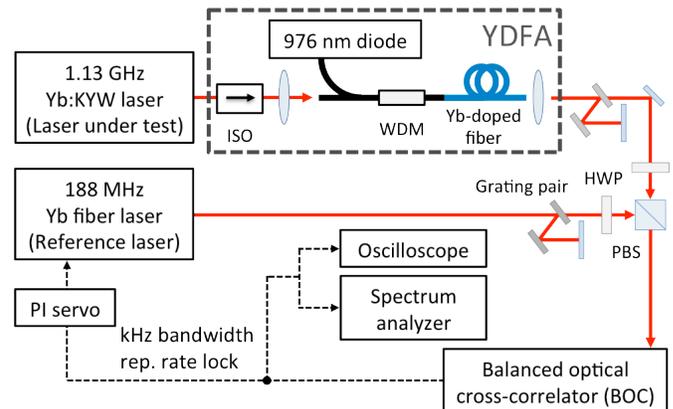

Fig. 2. Schematic of timing jitter characterization setup of a 1.13-GHz Yb:KYW mode-locked laser. HWP, half-wave plate; ISO, isolator; PBS, polarization beam splitter; PI, proportional-integral; WDM, wavelength-division multiplexing coupler; YDFA, Yb-doped fiber amplifier.

Curve (a) in Fig. 3 shows the measured timing jitter spectral density of the 1.13-GHz Yb:KYW laser. The measured jitter level is much higher than that of the Yb-fiber laser (curve (b), taken from the result in [12]), indicating that the current measurement is limited by the GHz Yb:KYW laser. Above ~300-kHz offset frequency, the measurement was limited at $10^{-8}$ fs$^2$/Hz level set by the noise floor of the used RF signal analyzer (curve (c)). This instrument noise floor sets the achievable resolution of our jitter spectral density measurement, which is ~0.41 fs in 10-MHz measurement bandwidth. Curve (d) shows the projected shot noise-limited detection sensitivity of the BOC, which is ~10 dB lower than the instrument limit (curve (c)). From 30 kHz to 300 kHz, the jitter spectrum follows the 1/f$^2$ (-20 dB/dec) slope (curve (e)), which comes from the quantum-limited timing jitter, as will be explained more in detail later with Fig. 4. The origin of higher slope in <30 kHz is not clear, but we believe that it might be caused by the combined effects of the influence of the repetition-rate locking loop (e.g., insufficient phase margin near the locking bandwidth) and technical noise introduced to the LUT (e.g., broadband acoustic noise and vibration). The minimal 3-dB locking bandwidth we could set is ~17.5 kHz, and the rms timing jitter of the free-running Yb:KYW laser integrated from 17.5 kHz to 10 MHz offset frequency is 0.70 fs.

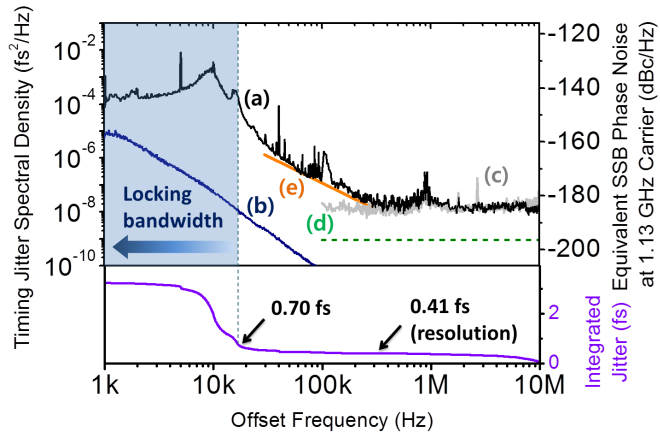

Fig. 3. Timing jitter measurement result. (a) Measured timing jitter of the 1.13-GHz Yb:KYW laser. (b) Measured timing jitter of the 188-MHz Yb-fiber laser (taken from [12]). (c) Measured noise floor of the RF signal analyzer used. (d) Shot-noise-limited noise floor of the BOC. (e) 1/f$^2$ (-20 dB/dec) slope.

We further analyzed the origins of the measured timing jitter using measured and known laser parameters, as shown in Fig. 4. The used measured and known laser parameters are following: intra-cavity pulse energy of 8 nJ, FWHM pulse width of 282 fs, net cavity dispersion of -1200 fs$^2$, intra-cavity loss of 3.3 %, center wavelength of 1037 nm, and SWCNT saturation fluence of 10 µJ/cm$^2$. We assumed the gain bandwidth of 16 nm [25] and the nonlinear refractive index of n$_2$ = 8.7×10$^{-16}$ cm$^2$/W [26] for the Yb:KYW crystal. By using these laser parameters, we calculated the nonlinear phase shift of 0.08 rad per cavity round-trip. We also used the measured RIN spectrum, as shown in Fig. 1(d), for the RIN-coupled jitter calculation.

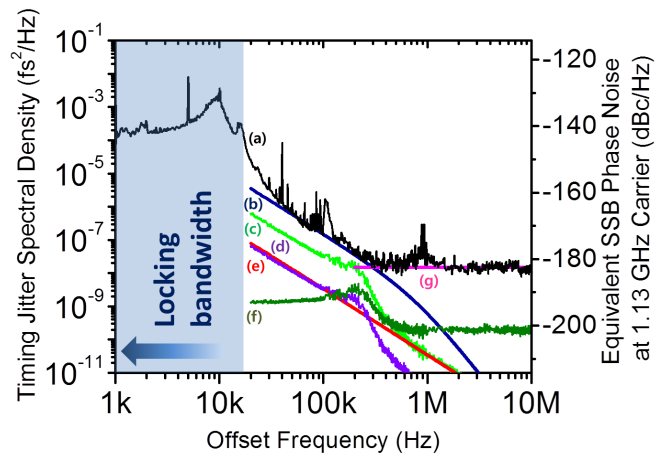

Fig. 4. Analysis of timing jitter sources. (a) Measured timing jitter spectrum. Calculated jitter spectra based on theory in [27,28]: (b) Gordon-Haus jitter (assuming excess noise factor of 18), (c) jitter from self-steepening (assuming 0.08 rad nonlinear phase shift per round-trip), (d) RIN-coupled jitter by a slow saturable absorber, (e) quantum-limited jitter directly from the ASE noise (assuming excess noise factor of 18), and (f) RIN-coupled jitter by the Kramers-Kronig relationship. (g) Measurement noise floor set by the used instrument.

In Fig. 4, we calculated and plotted projected timing jitter spectra originated from different effects, obtained from well-established analytic theories shown in [27,28]. When setting the excess noise factor (spontaneous emission factor) to 18, the measured result (curve (a)) can be fit with the projected Gordon-Haus jitter (curve (b)), i.e., the quantum-limited timing jitter originated from the center frequency fluctuation coupled by the intra-cavity dispersion, in the 30 kHz – 300 kHz range. Other curves indicate (c) timing jitter from self-steepening effect, (d) RIN-coupled jitter by the slow saturable absorber (SWCNT in this work), (e) quantum-limited timing jitter directly coupled from the ASE noise, and (f) RIN-coupled jitter by the Kramers-Kronig relationship. Curve (g) indicates the measurement limit set by the used signal analyzer, as explained in Fig. 3. Note that the quantum-limited jitter contributions (curves (b) and (e)) are multiplied by the fitted excess noise factor, whereas the RIN-originated jitter contributions (curves (c), (d) and (f)) are not. As the excess noise factor (which influences the Gordon-Haus jitter magnitude, curve (b)) and the nonlinear phase shift (which influences the self-steepening jitter magnitude, curve (c)) are not directly measured values and can be quite different from our prediction, it is difficult to clearly conclude which effect dominates the measured jitter performance. For example, if the nonlinear phase shift were just twice of our theoretical calculation, the self-steepening can be the dominant contributor to the measured jitter level. To exactly identify the origin, it would be necessary to resolve higher offset frequency above 300 kHz, where the Gordon-Haus jitter and self-steepening induced jitter has quite different spectral shape. Currently, we are limited by the achievable measurement resolution (curve (g)), and will need higher-resolution measurement in the future.

Provided the jitter measurement is not limited by the BOC resolution, the real jitter performance is expected to follow the combined effects of the Gordon-Haus jitter (curve (b) in Fig. 4) for <1.5 MHz offset frequency and the RIN-coupled jitter by the Kramers-Kronig relationship (curve (f) in Fig. 4) for >1.5 MHz offset frequency. Further assuming the RIN level is flat at -130 dB/Hz up to 565 MHz offset frequency, which is a reasonable assumption given the long (>μs) lifetime of a solid-state gain medium, the projected rms timing jitter integrated from 10 kHz to the full Nyquist frequency (565 MHz) is 0.50 fs. We believe that more engineering of the laser, such as lowering intra-cavity loss, increasing the pulse energy and shortening the pulse width, can further scale down the achievable timing jitter of the GHz pulse train. For example, if we use the laser parameters in [23], the projected rms timing jitter is 0.42 fs integrated up to the Nyquist frequency.

In summary, we demonstrated that a 1.13-GHz repetition rate optical pulse train with 0.70 fs high-frequency timing jitter (integration bandwidth of 17.5 kHz – 10 MHz, where the measurement instrument-limited noise floor contributes 0.41 fs in 10-MHz bandwidth) can be directly generated from a free-running, single-mode diode-pumped Yb:KYW laser mode-locked by SWCNT-coated saturable absorber mirrors. We also identified the major contributors to this jitter performance is the Gordon-Haus jitter and the timing jitter from self-steepening effect. As the jitter performance is not limited by technical noise sources originated from a slow saturable absorber, our result indicates that other types of diode-pumped GHz repetition rate solid-state lasers with different gain media, center wavelengths, and saturable absorbers may also observe similar sub-fs jitter performances, which will open up many opportunities for low-cost, compact, and self-starting GHz solid-state lasers for high-precision photonic signal processing applications in the near future.

This research was supported in part by the National Research Foundation (NRF) of Korea (under grants 2012R1A2A2A01005544 and 2013M1A3A3A02042273) and the Korea Research Institute of Standards and Science (KRISS, under grant 13011001). G. -H. K. acknowledges support from the Seoul Metropolitan Government of Korea (WR100001). S. Y. C. and F. R. acknowledge supports from the NRF of Korea (2008-0061906 and 2011-0017494).

### References

1. G. C. Valley, Opt. Express **15**, 1955 (2007).
2. J. Kim, M. J. Park, M. H. Perrott, and F. Kärtner, Opt. Express **16**, 16509 (2008).
3. A. Khilo, S. J. Spector, M. E. Grein, A. H. Nejadmalayeri, C. W. Holzwarth, M. Y. Sander, M. S. Dahlem, M. Y. Peng, M. W. Geis, N. A. DiLello, J. U. Yoon, A. Motamedi, J. S. Orcutt, J. P. Wang, C. M. Sorace-Agaskar, M. A. Popović, J. Sun, G. R. Zhou, H. Byun, J. Chen, J. L. Hoyt, H. I. Smith, R. J. Ram, M. Perrott, T. M. Lyszczarz, E. P. Ippen, and F. X. Kärtner, Opt. Express **20**, 4454 (2012).
4. A. O. J. Wiberg, Z. Tong, L. Liu, J. L. Ponsetto, V. Ataie, E. Myslivets, N. Alic, and S. Radic, Optical Fiber Communication Conference (OFC, 2012), OW3C.2
5. G. A. Keeler, B. E. Nelson, D. Agarwal, C. Debaes, N. C. Helman, A. Bhatnagar, and D. A. B. Miller, IEEE J. Sel. Top. Quantum Electron. **9**, 477 (2003).
6. Y. Song, C. Kim, K. Jung, H. Kim, and J. Kim, Opt. Express **19**, 14518 (2011).
7. T. K. Kim, Y. Song, K. Jung, C. Kim, H. Kim, C. H. Nam, and J. Kim, Opt. Lett. **36**, 4443 (2011).
8. A. J. Benedick, J. G. Fujimoto, and F. X. Kärtner, Nat. Photonics **6**, 97 (2012).
9. D. Li, U. Demirbas, A. Benedick, A. Sennaroglu, J. G. Fujimoto, and F. X. Kärtner, Opt. Express **20**, 23422 (2012).
10. E. Portuondo-Campa, R. Paschotta, and S. Lecomte, Opt. Lett. **38**, 2650 (2013).
11. C. Kim, S. Bae, K. Kieu, and J. Kim, Opt. Express **21**, 26533 (2013).
12. H. Kim, Y. Song, P. Qin, J. Shin, C. Kim, K. Jung, C. Wang, and J. Kim, Conference on Lasers and Electro Optics (CLEO, 2013), Paper CTh4M.4.
13. J. Chen, J. W. Sickler, P. Fendel, E. P. Ippen, F. X. Kärtner, T. Wilken, R. Holzwarth, and T. W. Hänsch, Opt. Lett. **33**, 959 (2008).
14. I. Ozdur, M. Akbulut, N. Hoghooghi, D. Mandridis, S. Ozharar, F. Quinlan, and P. J. Delfyett, IEEE Photon. Technol. Lett. **22**, 431 (2010).
15. D. Panasenko, P. Polynkin, A. Polynkin, J. V. Moloney, M. Mansuripur, and N. Peyghambarian, IEEE Photon. Technol. Lett. **18**, 853 (2006).
16. S. Zhou, D. G. Ouzounov, and F. W. Wise, Opt. Lett. **31**, 1041 (2006).
17. C. Lecaplain and P. Grelu, Opt. Express **21**, 10897 (2013).
18. L. Hou, M. Haji, J. Akbar, B. Qiu, and A. C. Bryce, Opt. Lett. **36**, 966 (2011).
19. H. Byun, M. Y. Sander, A. Motamedi, H. Shen, G. S. Petrich, L. A. Kolodziejski, E. P. Ippen, and F. X. Kärtner, Appl. Opt. **49**, 5577 (2010).
20. A. Schlatter, B. Rudin, S. C. Zeller, R. Paschotta, G. J. Spühler, L. Krainer, N. Haverkamp, H. R. Telle, and U. Keller, Opt. Lett. **30**, 1536 (2005).
21. J. B. Schlager, B. E. Callicoatt, R. P. Mirin, N. A. Sanford, D. J. Jones, and J. Ye, Opt. Lett. **28**, 2411 (2003).
22. V. J. Wittwer, C. A. Zaugg, W. P. Pallmann, A. E. H. Oehler, B. Rudin, M. Hoffmann, M. Golling, Y. Barbarin, T. Südmeyer, and U. Keller, IEEE Photon. J. **3**, 658 (2011).
23. H. Yang, C. Kim, S. Y. Choi, G. Kim, Y. Kobayashi, F. Rotermund, and J. Kim, Opt. Express **20**, 29518 (2012).
24. J. Kim, J. Chen, J. Cox, and F. X. Kärtner, Opt. Lett. **32**, 3519 (2007).
25. N. V. Kuleshov, A. A. Lagatsky, A. V. Podlipensky, V. P. Mikhailov, and G. Huber, Opt. Lett. **22**, 1317 (1997).
26. K. V. Yumashev, N. N. Posnov, P. V. Prokoshin, V. L. Kalashnikov, F. Mejid, I. G. Poloyko, V. P. Mikhailov, and V. P. Kozich, Opt. Quantum Electron. **32**, 43 (2000).
27. R. Paschotta, Appl. Phys. B **79**, 163 (2004).
28. R. Paschotta, Opt. Express **18**, 5041 (2010).

Full references (with titles)

1. G. C. Valley, "Photonic analog-to-digital converters," Opt. Express **15**, 1955-1982 (2007).
2. J. Kim, M. J. Park, M. H. Perrott, and F. Kärtner, "Photonic subsampling analog-to-digital conversion of microwave signals at 40-GHz with higher than 7-ENOB resolution," Opt. Express **16**, 16509-16515 (2008).
3. A. Khilo, S. J. Spector, M. E. Grein, A. H. Nejadmalayeri, C. W. Holzwarth, M. Y. Sander, M. S. Dahlem, M. Y. Peng, M. W. Geis, N. A. DiLello, J. U. Yoon, A. Motamedi, J. S. Orcutt, J. P. Wang, C. M. Sorace-Agaskar, M. A. Popović, J. Sun, G. R. Zhou, H. Byun, J. Chen, J. L. Hoyt, H. I. Smith, R. J. Ram, M. Perrott, T. M. Lyszczarz, E. P. Ippen, and F. X. Kärtner, "Photonic ADC: overcoming the bottleneck of electronic jitter," Opt. Express **20**, 4454–4469 (2012).
4. A. O. J. Wiberg, Z. Tong, L. Liu, J. L. Ponsetto, V. Ataie, E. Myslivets, N. Alic, S. Radic, "Demonstration of 40 GHz analog-to-digital conversion using copy-and-sample-all parametric processing," Optical Fiber Communication Conference (OFC, 2012), OW3C.2
5. G. A. Keeler, B. E. Nelson, D. Agarwal, C. Debaes, N. C. Helman, A. Bhatnagar, D. A. B. Miller, " The benefits of ultrashort optical pulses in optically interconnected systems," IEEE J. Sel. Top. Quantum Electron. **9**, 477-485 (2003).
6. A. J. Benedick, J. G. Fujimoto, and F. X. Kärtner, "Optical flywheels with attosecond jitter," Nat. Photonics **6**, 97-100 (2012).
7. D. Li, U. Demirbas, A. Benedick, A. Sennaroglu, J. G. Fujimoto, and F. X. Kärtner, "Attosecond timing jitter pulse trains from semiconductor saturable absorber mode-locked Cr:LiSAF lasers," Opt. Express **20**, 23422-23435 (2012).
8. E. Portuondo-Campa, R. Paschotta, and S. Lecomte, "Sub-100 attosecond timing jitter from low-noise passively mode-locked solid-state laser at telecom wavelength," Opt. Lett. **38**, 2650-2653 (2013).
9. Y. Song, C. Kim, K. Jung, H. Kim, and J. Kim, "Timing jitter optimization of mode-locked Yb-fiber lasers toward the attosecond regime," Opt. Express **19**, 14518-14525 (2011).
10. T. K. Kim, Y. Song, K. Jung, C. Kim, H. Kim, C. H. Nam, and J. Kim, "Sub-100-as timing jitter optical pulse trains from mode-locked Er-fiber lasers," Opt. Lett. **36**, 4443-4445 (2011).
11. C. Kim, S. Bae, K. Kieu, J. Kim, "Sub-femtosecond timing jitter, all-fiber, CNT-mode-locked Er-laser at telecom wavelength," Opt. Express **21**, 26533-26541 (2013).
12. H. Kim, Y. Song, P. Qin, J. Shin, C. Kim, K. Jung, C. Wang, and J. Kim, "Reduction of timing jitter to the sub 20-attosecond regime in free-running femtosecond mode-locked fiber lasers," Conference on Lasers and Electro Optics (CLEO, 2013), CTh4M.4
13. J. Chen, J. W. Sickler, P. Fendel, E. P. Ippen, F. X. Kärtner, T. Wilken, R. Holzwarth, and T. W. Hänsch, "Generation of low-timing-jitter femtosecond pulse trains with 2 GHz repetition rate via external repetition rate multiplication," Opt. Lett. **33**, 959-961 (2008).
14. I. Ozdur, M. Akbulut, N. Hoghooghi, D. Mandridis, S. Ozharar, F. Quinlan, and P. J. Delfyett, "A Semiconductor-Based 10-GHz Optical Comb Source With Sub 3-fs Shot-Noise-Limited Timing Jitter and ~500-Hz Comb Linewidth," IEEE Photon. Technol. Lett. **22**, 431-433 (2010).
15. D. Panasenko, P. Polynkin, A. Polynkin, J. V. Moloney, M. Mansuripur, and N. Peyghambarian, "Er-Yb Femtosecond Ring Fiber Oscillator With 1.1-W Average Power and GHz Repetition Rates," IEEE Photon. Technol. Lett. **18**, 853-855 (2006).
16. S. Zhou, D. G. Ouzounov, F. W. Wise, "Passive harmonic mode-locking of a soliton Yb fiber laser at repetition rates to 1.5 GHz," Opt. Lett. **31**, 1041-1043 (2006).
17. C. Lecaplain and P. Grelu, "Multi-gigahertz repetition-rate-selectable passive harmonic mode locking of a fiber laser," Opt. Express **21**, 10897-10902 (2013).
18. L. Hou, M. Haji, J. Akbar, B. Qiu, and A. C. Bryce, "Low divergence angle and low jitter 40 GHz AlGaInAs/InP 1.55 μm mode-locked lasers," Opt. Lett. **36**, 966-968 (2011).
19. H. Byun, M. Y. Sander, A. Motamedi, H. Shen, G. S. Petrich, L. A. Kolodziejski, E. P. Ippen, and F. X. Kärtner, "Compact, stable 1 GHz femtosecond Er-doped fiber lasers," Appl. Opt. **49**, 5577-5582 (2010).
20. A. Schlatter, B. Rudin, S. C. Zeller, R. Paschotta, G. J. Spühler, L. Krainer, N. Haverkamp, H. R. Telle and U. Keller, "Nearly quantum-noise-limited timing jitter from miniature Er:Yb:glass lasers," Opt. Lett. **30**, 1536-1538 (2005).
21. J. B. Schlager, B. E. Callicoatt, R. P. Mirin, N. A. Sanford, D. J. Jones, and J. Ye, "Passively mode-locked glass waveguide laser with 14-fs timing jitter," Opt. Lett. **28**, 2411-2413 (2003).
22. V. J. Wittwer, C. A. Zaugg, W. P. Pallmann, A. E. H. Oehler, B. Rudin, M. Hoffmann, M. Golling, Y. Barbarin, T. Südmeyer, and U. Keller, "Timing Jitter Characterization of a Free-Running SESAM Mode-locked VECSEL," IEEE Photon. J. **3**, 658-664 (2011).
23. H. Yang, C. Kim, S. Y. Choi, G. Kim, Y. Kobayashi, F. Rotermund, and J. Kim, "1.2-GHz repetition rate, diode-pumped femtosecond Yb:KYW laser mode-locked by a carbon nanotube saturable absorber mirror," Opt. Express **20**, 29518-29523 (2012).
24. J. Kim, J. Chen, J. Cox, and F. X. Kärtner, "Attosecond-resolution timing jitter characterization of free-running mode-locked laser," Opt. Lett. **32**, 3519-3521 (2007).
25. N. V. Kuleshov, A. A. Lagatsky, A. V. Podlipensky, V. P. Mikhailov, and G. Huber, "Pulsed laser operation of Yb-doped KY(WO$_4$)$_2$ and KGd(WO$_4$)$_2$," Opt. Lett. **22**, 1317-1319 (1997).
26. K. V. Yumashev, N. N. Posnov, P. V. Prokoshin, V. L. Kalashnikov, F. Mejid, I. G. Poloyko, V. P. Mikhailov, and V. P. Kozich, "Z-scan measurements of nonlinear refraction and Kerr-lens mode-locking with Yb$^{3+}$:KY(WO$_4$)$_2$", Opt. Quantum Electron. **32**, 43-48 (2000).
27. R. Paschotta, "Noise of mode-locked lasers (Part II): timing jitter and other fluctuations," Appl. Phys. B **79**, 163-173 (2004).
28. R. Paschotta, "Timing jitter and phase noise of mode-locked fiber lasers," Opt. Express 18, 5041-5054 (2010).